\documentclass[letterpaper]{article} 
\usepackage[draft]{aaai25}  
\usepackage{times}  
\usepackage{helvet}  
\usepackage{courier}  
\usepackage[hyphens]{url}  
\usepackage{graphicx} 
\usepackage{natbib}  
\usepackage{caption} 
\usepackage{algorithm}

\usepackage{newfloat}
\usepackage{listings}
\DeclareCaptionStyle{ruled}{labelfont=normalfont,labelsep=colon,strut=off} 
\floatstyle{ruled}
\newfloat{listing}{tb}{lst}{}
\floatname{listing}{Listing}
\usepackage{amsmath, amssymb, amsthm, amsxtra}
\usepackage{algpseudocode}
\usepackage{comment}
\usepackage{url}
\usepackage{multirow}
\usepackage{appendix}
\usepackage{lipsum}
\usepackage{booktabs}
\usepackage{placeins}

\title{Advancing Marine Heatwave Forecasts: An Integrated Deep Learning Approach}
\author {
    Ding Ning\textsuperscript{\rm 1},
    Varvara Vetrova\textsuperscript{\rm 1},
    Yun Sing Koh\textsuperscript{\rm 2},
    Karin R. Bryan\textsuperscript{\rm 3}
}
\affiliations {
    \textsuperscript{\rm 1}School of Mathematics and Statistics, University of Canterbury, Christchurch, New Zealand\\
    \textsuperscript{\rm 2}School of Computer Science, University of Auckland, Auckland, New Zealand\\
    \textsuperscript{\rm 3}School of Environment, University of Auckland, Auckland, New Zealand\\
    ding.ning@pg.canterbury.ac.nz, varvara.vetrova@canterbury.ac.nz, \{y.koh,karin.bryan\}@auckland.ac.nz
}

\usepackage{bibentry}

\begin{document}

\maketitle

\begin{abstract}
Marine heatwaves (MHWs), an extreme climate phenomenon, pose significant challenges to marine ecosystems and industries, with their frequency and intensity increasing due to climate change. This study introduces an integrated deep learning approach to forecast short-to-long-term MHWs on a global scale. The approach combines graph representation for modeling spatial properties in climate data, imbalanced regression to handle skewed data distributions, and temporal diffusion to enhance forecast accuracy across various lead times. To the best of our knowledge, this is the first study that synthesizes three spatiotemporal anomaly methodologies to predict MHWs. Additionally, we introduce a method for constructing graphs that avoids isolated nodes and provide a new publicly available sea surface temperature anomaly graph dataset. We examine the trade-offs in the selection of loss functions and evaluation metrics for MHWs. We analyze spatial patterns in global MHW predictability by focusing on historical hotspots, and our approach demonstrates better performance compared to traditional numerical models in regions such as the middle south Pacific, equatorial Atlantic near Africa, south Atlantic, and high-latitude Indian Ocean. We highlight the potential of temporal diffusion to replace the conventional sliding window approach for long-term forecasts, achieving improved prediction up to six months in advance. These insights not only establish benchmarks for machine learning applications in MHW forecasting but also enhance understanding of general climate forecasting methodologies.
\end{abstract}

%

\section{Introduction}

Marine heatwaves (MHWs) are observed across the world and have strong effects on marine ecosystems. The devastating impact on marine ecosystems caused by MHWs can bring irreversible loss of species or foundation habitats \citep{oliver2019projected}, for example, mass coral bleaching and substantial declines in kelp forests and seagrass meadows \citep{holbrook2020keeping}. MHWs also affect aquaculture businesses and area-restricted fisheries due to the change in the distribution of sea life with follow-on effects on production \citep{hobday2018framework}, such as in mussel, oyster and salmon farms. Anthropogenic climate change is expected to cause an increase in both the intensity and frequency of MHWs \citep{hobday2016hierarchical}.

The variability of sea surface temperatures (SSTs), or SST anomalies (SSTAs), is a fundamental variable in defining and predicting MHWs. SSTAs are the departures from the average sea surface temperature over a certain period \citep{ham2019deep,pravallika2022prediction,taylor2022deep}. Following \cite{hobday2016hierarchical}, we define monthly MHWs as events where the SSTAs are greater than the 90\textsuperscript{th} percentile of the SSTAs based on a monthly climatology.  For example, a MHW in January 2024 at a particular location would be defined as present if an average SSTA for January 2024 is greater than 90\% of January SSTAs over a specified period (climatology) at that location.

The ability to accurately predict MHWs, three months in advance, for instance, would enable effective mitigation strategies to reduce their adverse impacts. This research proposes an integrated deep learning (DL) framework for global monthly MHW forecasts, which combines graph representation, imbalanced regression, and temporal diffusion. We evaluate the data-driven predictability of MHWs using the selected dataset and DL approach, focusing on both spatial and temporal dimensions, and comparing it with existing numerical models \citep{jacox2022global}.

\textbf{Contributions.} Our focus is on integrating DL techniques for forecasting MHWs on both global and selected regional scales, across monthly to seasonal timescales. The main contributions of this paper can be summarized as follows:
\begin{itemize}
    \item We propose a DL framework that combines three methodologies: graph representation, imbalanced regression, and temporal diffusion, to enhance the forecasting of MHWs from short to long lead times on a global scale. To our knowledge, this is the first study to synergistically combine these methodologies to predict MHWs. Furthermore, the framework allows for potential application to other climate extreme forecasts. Each component's contribution is detailed below:
    \item We present a new graph construction approach designed to overcome an issue with node isolation and generate a new publicly available SSTA graph dataset. Generally, the approach has improved MHW predictions both on average and at 12 selected MHW hotspot locations.
    \item We optimize loss functions tailored for MHW prediction. We find that imbalanced regression loss functions notably increased the detection of MHW occurrences, while the standard MSE loss tended to lower the false alarm rate. Our approach achieved effective average symmetric extremal dependence index (SEDI) scores, and at the selected MHW hotspots, the imbalanced regression loss functions generally outperformed the standard MSE.
    \item We introduce an adaptation of the temporal diffusion process that eliminates the need for sliding windows conventionally for climate forecasting. We find that the diffusion method reduces input requirements and enhances long-term MHW predictive skills. To the best of our knowledge, this is the first attempt to combine temporal diffusion with graph neural networks for climate forecasts. 
    \item For spatial outcomes, our MHW models, evaluated by the SEDI, generated forecasts with distinct spatial patterns compared to numerical models on a global scale. Temporally, evaluated by combining the SEDI and the critical success index (CSI), MHWs could be forecasted up to six months in advance. The DL approach is also expected to enhance inference speed compared to numerical models.
\end{itemize}

\section{Related Work}

This section reviews the existing numerical and DL models used in MHW forecasts and the relevant DL methods.

\subsection{Marine Heatwave Forecasts}

\textbf{Numerical models.} Physics-based numerical models utilize fundamental laws of physics to simulate oceanic and atmospheric processes that lead to MHWs to make forecasts, widely used in the dynamical prediction systems \citep{merryfield2013canadian,saha2014ncep,vecchi2014seasonal}. For example, a case study of the California Current System MHW of 2014-2016 used eight global coupled climate prediction systems to predict the MHW up to eight months ahead \citep{jacox2019predicting}. In this case, two of the four phases were predicted well by dynamic models but two others were missed. Moreover, numerical dynamic models are resource intensive, making it difficult to run enough simulations to perform probabilistic projections. At the same time, there have been more and more observations and completed hindcasts, so statistical explorations of the existing data to make forecasts might be a useful alternative to dynamic modeling. \cite{jacox2022global} provided a comprehensive examination of numerical MHW forecasts on a global scale, from one to 12 months in advance, using the OISST v.2.1 dataset \citep{banzon2016long,reynolds2007daily}. The main evaluation metric was the SEDI \citep{ferro2011extremal}. The results showed that the forecasts deteriorated as the number of lead months increased, and spatially, better forecasts were observed in the equatorial Pacific (the ENSO zone), in the equatorial Indian and Arabian Sea, around Southeast Asia, in the north Tasman Sea, near the northwest coast of North America, in the southeast Pacific next to Antarctica, and in the equatorial Atlantic and Caribbean Sea.

\textbf{Deep learning models.} DL models for MHWs initially focused on predicting SSTAs, with the aim of forecasting climate oscillations such as the El Ni\~{n}o-Southern Oscillation (ENSO), the Indian Ocean Dipole (IOD) and climate extreme events. For example, \cite{ham2019deep,ham2021unified} used a CNN with SSTAs as one of the inputs to forecast the seasonal Nino3.4 SSTA index up to 18 months ahead, which is a climate index based on the SST field over the tropical eastern Pacific that characterizes the El Ni\~{n}o-Southern Oscillation (ENSO) in the equatorial Pacific Ocean. \cite{ratnam2020machine} proposed a fully-connected neural network (FCN) to forecast SSTAs over the Indian Ocean as indicators of the IOD oscillations. Also, the IOD forecasts have been made using an LSTM \citep{pravallika2022prediction} and a CNN \citep{fengpredictability}. For MHWs, a CNN was developed to forecast SSTs and MHWs around Australia \citep{boschetti2022sea}. Later work started to explore combining multiple neural network classes. \cite{taylor2022deep} integrated a U-Net and an LSTM with the sliding window method to forecast SSTs up to 24 months ahead, validated with specific SST variability-related oscillations (ENSO and IOD) and climate events (the Blob marine heatwave off the west coast of the United States). The U-Net-LSTM predicted SSTs and then converted SSTs into the Nino3.4 and Nino4 indices and the Blob index. Another recent study explored the potential of graph representation to contribute to global SST and SSTA predictions and demonstrated another pathway for MHW forecasts \citep{ning2024harnessing}. The work showed usable up-to-two-year-ahead SST prediction with a combination of a GraphSAGE model \citep{hamilton2017inductive} and a recursive model, and one-month-ahead SSTA prediction using another GraphSAGE with further analysis of the global spatial pattern of predicted SSTAs and SSTAs at 12 selected historical MHW hotspots \citep{oliver2021marine}. The model produced a better average RMSE than the persistence model for one-month-ahead SSTA forecasting and revealed the spatial pattern of the forecast made by the graph approach. For temporal pattern learning, the study used the sliding window method when training the GraphSAGE and the recursive approach for making long-term forecasting using short-term prediction. The results showed that the long-term SST prediction was largely dependent on the forecasting window length, which is a trade-off between the amount of input data and long-term predictive performance. Also, the recursive model could produce usable long-term SST forecasts due to the seasonality inherent in SSTs, but failed for SSTAs after seasonality was removed. Moreover, computing SSTAs with the predicted SSTs yielded poorer results than persistence \citep{ning2024harnessing}.

\subsection{Deep Learning Methods}

\textbf{Graph representation and GraphSAGE.} Previous studies have detailed that graph representation is well structured to take advantage of climatological data at a global scale \citep{ning2024harnessing}. The teleconnections of climate events, driven by atmospheric and oceanic circulation or large-scale Rossby waves \citep{tsonis2006networks}, have been an important factor for DL modeling \citep{cachay2021world,taylor2022deep} and can be effectively captured using graph structures. Moreover, traditional grids and CNNs have inherent challenges such as handling missing values, lack of rotation equivariance \citep{defferrard2019deepsphere}, and computational receptive field issues \citep{luo2016understanding} for Earth data. GraphSAGE provides a scalable approach for embedding nodes in large graphs and, with its inductive learning capabilities, is suitable for dynamic graphs. GraphSAGE offers an LSTM aggregator that, unlike the mean and pooling aggregators, exploits the sequential dependencies among node features \citep*{hamilton2017inductive}. Despite this, additional approaches capable of handling temporal patterns and integrating with GraphSAGE are still desirable.

\textbf{Temporal diffusion.} The diffusion process effectively models the distribution of the data, and has been widely applied to the vision domain \citep{ho2022imagen,yang2023diffusion,esser2023structure,croitoru2023diffusion}. Apart from learning visual and spatial patterns, diffusion could be used to model temporal patterns. A dynamics-informed diffusion framework was proposed for efficient training of diffusion models tailored for probabilistic spatiotemporal forecasting \citep{cachay2023dyffusion}. One main component within the framework is the temporal diffusion mechanism, which leverages temporal dynamics within data, integrating these dynamics directly using diffusion steps. This mechanism achieved more accurate and computationally efficient multi-step and long-range forecasting \citep{cachay2023dyffusion}. In the weather and climate domains, temporal diffusion has been applied using U-Nets or U-Net-like autoencoders for forecasting 500 hPa geopotential heights and 850 hPa temperatures \citep{hua2024weather}, as well as two-meter air temperatures and wind states \citep{watt2024generative}. However, combining temporal diffusion with graph representation has not been attempted before, which is a key focus of this study.

\textbf{Imbalanced regression.} In terms of DL for SSTA-related oscillation or event forecasts, the majority of the attempts have focused on reproducing the average conditions but not the extremes, such as the ENSO \citep{ham2019deep,taylor2022deep,boschetti2022sea} and IOD \citep{wu2019seasonal,ratnam2020machine,fengpredictability} forecasts. These models generally formulated the learning tasks for SSTA forecasts as a standard regression or classification task (using a standard loss function). Nevertheless, the extremes like MHWs are much more damaging to ecosystems and productivity, and their forecasts are useful for the climate domain. As SSTAs are a continuous variable, the forecasting tasks could be viewed as imbalanced regression, enabling the exploration of techniques to handle the SSTA extremes associated with MHWs. Approaches to tackle imbalanced regression are generally categorized into two groups: the data-based where data are re-sampled \citep{torgo2013smote,branco2017smogn,yang2021delving}, and the model-based in which the loss function is re-weighting or adjusting to compensate for label imbalance \citep{yang2021delving,ren2022balanced}.

This study aims to address MHW forecasting challenges by proposing a DL framework that combines graph representation, imbalanced regression, and temporal diffusion.

\section{Data}

\textbf{Dataset.} The dataset for MHW forecasts was extracted from ERA5 \citep{hersbach2020era5}. Specifics on data extraction and characteristics are detailed in the studies \citep{taylor2022deep,ning2024harnessing}. The extracted dataset comprises SST measurements, encompassing a spatial extent of [720, 1440] pixels along latitudes and longitudes and spanning from January 1940 to December 2022, with a temporal resolution of one month, totaling 996 months.

\textbf{Data preprocessing.} Following \cite{taylor2022deep,ning2024harnessing}, we preprocessed the extracted SST data to a smaller [64, 128, 996] latitude (64°S to 62°N in 2° increments), longitude (180°W to 180°E in 2.8125° increments), and monthly-resolution grid. The unit of SSTs is Kelvin. We divided the 3D data into training and test sets based on their temporal attributes. Specifically, 840 2D grids from time steps starting in January 1940 served as training features, while the remaining 2D grids constituted the test features. The window lag technique was employed to extract the corresponding labels, and the number of test grids varied depending on the window size.

\section{Methodology}

\textbf{Graph construction.} Following \cite{ning2024harnessing}, we constructed the SSTA graphs from the 2D SST grids by defining node features and adjacencies, by setting a threshold $c$ for the Kendall rank correlation coefficient to evaluate the similarity between the features of every pair of nodes, where node pairs with a correlation above the threshold are considered related. However, due to the fixed threshold $c$, some nodes had few or no connections to other nodes. Then we proposed another approach: the Kendall rank correlation coefficients between every two nodes were computed first, and then we assigned a fixed number $m \in \mathbb{N}$ that every node has no less than $m$ edges to ensure no isolated nodes, and the average node degree is $2m$. The coefficients for every node were sorted, and the edges with the highest coefficients were kept for each node. The adjacency matrices for the graph sets are summarized in Table \ref{tb:adjmat}. The steps for computing and normalizing SSTAs followed \cite{ning2024harnessing}. Accessibility to the new SSTA graph dataset can be found in the Appendix.
\begin{algorithm}[h!]
  \caption{Sorted Kendall rank correlation-based adjacency matrix construction.}
  \label{adjmat1}
      \begin{algorithmic}[1]
        \State \textbf{Input:} Data matrix $\mathbf{X} \in \mathbb{R}^{n \times d}$, parameter $m$
        \State \textbf{Output:} Adjacency matrix $\mathbf{A} \in \mathbb{R}^{n \times n}$

        \State $\text{adj\_list} \gets \emptyset$

        \For{$i \in \{1, 2, \ldots, n\}$}
          \State $\text{correlations} \gets \emptyset$
          \For{$j \in \{1, 2, \ldots, n\} \setminus \{i\}$}
            \State $\tau_{ij} \gets \text{kendall\_tau}(\mathbf{X}_i, \mathbf{X}_j)$
            \State $\text{correlations} \gets \text{correlations} \cup \{(j, \tau_{ij})\}$
          \EndFor
          \State $\text{correlations} \gets \text{sort}(\text{correlations}, \text{by } -|\tau_{ij}|)[:m]$
          \For{$(j, \_) \in \text{correlations}$}
            \State $\text{adj\_list} \gets \text{adj\_list} \cup \{(i, j), (j, i)\}$
          \EndFor
        \EndFor

        \State $\text{adj\_list} \gets \text{sort}(\text{adj\_list})$
        \State $\mathbf{A} \gets \text{convert}(\text{adj\_list}, \text{to 2D matrix})$
        \State \Return $\mathbf{A}$
      \end{algorithmic}
\end{algorithm}

\begin{table*}[h!]
\centering \small
\begin{tabular}{l l l l}
\toprule
Method for defining the adjacency matrix & Average node degree & Max node degree & Min node degree \\ \midrule
$c=0.8$                               & 33.95               & 315             & 0               \\
$c=0.75$                              & 130.75              & 989             & 0               \\
$m=25$                                & 50.00               & 101             & 25              \\
$m=50$                                & 100.00              & 212             & 50              \\
$c=0.8(12)$                           & 57.74               & 5773            & 12              \\
$m=25(12)$                            & 74.00               & 5773            & 37              \\ \bottomrule
\end{tabular}
\caption{Adjacency matrices for constructing graphs. The graph sets with ``12'' in parentheses indicate we added full connectivity to the 12 MHW hotspot nodes with all the other nodes.}
\label{tb:adjmat}
\end{table*}

\textbf{Graph neural network.} We used GraphSAGE as the basic architecture. We experimented with the mean, pooling, and LSTM aggregators, using the mean as the default. The GraphSAGE model is detailed in the Appendix.

\textbf{Imbalanced loss functions.} We selected three loss functions as the objective function for training neural network models. The mean squared error (MSE) is commonly used for normal regression tasks and the baseline loss function used to compare the other two loss functions for imbalanced regression. The second loss function is the balanced MSE (BMSE) \citep{ren2022balanced}. In our experiments, we used a fixed noise variable value $\sigma=0.02$, where the selection of hyperparameter values was verified through our exploratory experiments. For MHW forecasts, we customized a weighted MSE (WMSE) for imbalanced regression as the third loss function. Using the definition of MHWs, in the training process, we assigned a larger weight $w>1$ to the mean square errors for the observations greater than the 90\textsuperscript{th} percentile and a weight of 1 otherwise. The larger weight emphasizes the importance of MHW predictions over non-MHW predictions. The weight for both types of observations is multiplied by a power function of the absolute values of the corresponding predictions, causing the weight to scale with the predictions. The WMSE is:
\begin{align}
&L(y,\hat{y})=\frac{1}{N}\sum_{i=1}^N l_i,\\
&l_i=\begin{cases}
(w\cdot |y_i|^{\alpha})(\hat{y}_i-y_i)^2;
& y_i>y_{90\%},\\
|y_i|^{\alpha}(\hat{y}_i-y_i)^2;
& otherwise,
\end{cases}
\end{align}
where $y_{90\%}$  is the 90\textsuperscript{th}  percentile of $y$, and $w$  is the controllable weight hyperparameter. $w\cdot |y|^{\alpha}$ is a scaling weight term. $\alpha$ and $w$ are the controllable hyperparameters, where $\alpha\in[1,+\infty]$ and $w\in[1,+\infty]$ are suggested. In this loss function, the weight scales with the observed values $y$ and $\alpha$ control the degree of scaling. The absolute value of $y$ ensures the term to be a real number. Similar to the standard weighted MSE, an additional weight $w$ is multiplied to separate the errors between MHW observations and predictions. In our experiments, we used fixed values $\alpha=2$ and $w=2$. The BMSE and WMSE loss functions were chosen based on the Friedman test  \citep{friedman1937use,friedman1940comparison} in earlier experiments (not included in this manuscript).

\textbf{Sliding window.} The sliding window method uses a subset of data points in the time series as inputs rather than all previous data points, and has been used in SST-related forecasts \citep{ham2019deep,taylor2022deep,ning2024harnessing}. The corresponding hyperparameter, window size, determines the number of previous time steps as predictors. \cite{ham2019deep} used a window size of 3 and \cite{taylor2022deep,ning2024harnessing} used 12 (one year). Our default window size was 12.

\begin{figure*}[h!]
  \centering
  \includegraphics[width=0.85\textwidth]{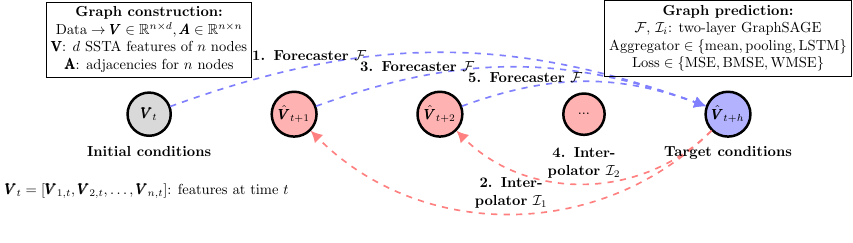}
  \caption{Overview of the diffusion-GraphSAGE with imbalanced losses for MHW forecasts.}\label{fg:process}
\end{figure*}

\textbf{Temporal diffusion.} We used a temporal diffusion process similar to the DYffusion model \citep{cachay2023dyffusion}. We propose the following modifications where the forecaster $\mathcal{F}$ and interpolator(s) $\mathcal{I}_i$ are analogous GraphSAGE models using the same configurations except two extensions. The first is that $\mathcal{I}_i$ has a dropout layer following each SAGE convolutional layer to introduce stochasticity, while $\mathcal{F}$ does not have dropout. The second is the output and input. For $\mathcal{F}$, the output is the prediction at the target long lead time step, and the input is windowed observations and interpolated prediction(s) between the last observation and the target prediction. For $\mathcal{I}_i$, the output is the prediction at one of the time steps between the last observation and the target prediction, and the input is the windowed observations and the target prediction. One or several interpolators could be trained, and the number of interpolators is the number of time steps between the last observation and the target prediction. For each epoch, $\mathcal{F}$ was trained at the beginning, followed by training one interpolator and refining $\mathcal{F}$ iteratively. The training process is detailed in the Appendix. We used the temporal diffusion for longer lead forecasts. Figure \ref{fg:process} provides a brief visualization of the whole process.

\textbf{Evaluation metrics.}  We used the precision, recall, CSI \citep{schaefer1990critical}, and SEDI to evaluate the prediction for MHWs. The CSI is equal to the total number of true positive predictions for extremes divided by the sum of the predictions for extremes plus the number of misses. We first used neural network models to predict SSTAs as a regression task. Then we categorized the predicted values by regressor into two classes: the MHW ($\hat{y}_i>y_{90\%}$) and the normal or non-MHW ($\hat{y}_i\leq y_{90\%}$). The SEDI has been used to evaluate numerical MHW forecasts \citep{jacox2022global} and is calculated as follows:
\begin{align}
\text{SEDI} = \frac{\log f - \log h - \log \left( 1 - f \right) + \log \left( 1 - h \right)}{\log f + \log h + \log \left( 1 - f \right) + \log \left( 1 - h \right)},
\end{align}
where $h$ is the hit rate (the ratio of true positives to total observed events), and $f$ is the false alarm rate (the ratio of false positives to total observed non-events). The maximum SEDI score is one, and scores above (below) zero indicate forecasts better (worse) than random chance \citep{jacox2022global}. The SEDI was the primary metric used to control early stops during model training.

\textbf{Reproducibility.} Details on reproducibility and code accessibility are provided in the Appendix.

\section{Experiments}

This sections details the experimental setups and methods used to evaluate the proposed models and approaches.

\subsection{Graph Construction and Loss Functions}

We tested combinations of three loss functions and two graph construction approaches, each with two sizes of adjacency matrices to examine the impact of the graph size (Table \ref{tb:loss_adj}). The four sets of configurations returned similar average CSIs and SEDIs. There is a trade-off between the average precision and recall. For short-term MHW forecasts, both the CSI and SEDI were appropriate evaluation metrics. Using the SEDI as the primary metric, the adjacency matrices $c=0.75$ and $m=25$ provided better performance, and in terms of loss functions, the BMSE and WMSE could produce better MHW forecasts. We also noticed that larger adjacency matrices did not necessarily return better results, with the effect varying between the threshold and ranking approaches. This possibly implies that some connected nodes through extra constructed edges did not provide useful information for feature aggregation and prediction.
\begin{table*}[h!]
\centering \small
\begin{tabular}{llllll}
\toprule
Adjacency matrix                   & Loss   & Precision   & Recall      & CSI         & SEDI        \\ \midrule
\multirow{3}{*}{$c=0.8$}  & MSE       & 0.591±0.14  & 0.7387±0.16 & 0.4949±0.14 & 0.6793±0.15 \\
                          & BMSE      & 0.5827±0.14 & 0.7509±0.16 & 0.4933±0.14 & 0.6791±0.15 \\
                          & WMSE      & 0.579±0.14  & 0.7558±0.15 & 0.4927±0.14 & 0.6793±0.15 \\ \midrule
\multirow{3}{*}{$c=0.75$} & MSE       & 0.5882±0.14 & 0.7446±0.16 & 0.4944±0.14 & 0.6796±0.15 \\
                          & BMSE      & 0.5885±0.14 & 0.7463±0.15 & 0.4965±0.14 & 0.6812±0.15 \\
                          & WMSE      & 0.5805±0.14 & 0.7569±0.16 & 0.4942±0.14 & 0.6805±0.15 \\ \midrule
\multirow{3}{*}{$m=25$}   & MSE       & 0.5916±0.14 & 0.7361±0.16 & 0.4924±0.14 & 0.677±0.15  \\
                          & BMSE      & 0.6016±0.14 & 0.7296±0.16 & 0.4982±0.14 & 0.6827±0.15 \\
                          & WMSE      & 0.5921±0.14 & 0.7408±0.16 & 0.4961±0.14 & 0.6821±0.15 \\ \midrule
\multirow{3}{*}{$m=50$}   & MSE       & 0.5974±0.14 & 0.7364±0.16 & 0.4981±0.14 & 0.6832±0.15 \\
                          & BMSE      & 0.586±0.15  & 0.7479±0.16 & 0.4903±0.14 & 0.6768±0.15 \\
                          & WMSE      & 0.5821±0.14 & 0.7501±0.16 & 0.4904±0.14 & 0.6759±0.15 \\ \bottomrule
\end{tabular}
\caption{Evaluation across all nodes for one-month-ahead forecasts of MHWs (precision, recall, CSI, and SEDI). The mean of each metric across all nodes is reported. The notation ``±'' indicates the standard deviation.}\label{tb:loss_adj}
\end{table*}

In the early stage of tuning hyperparameters, we initially used the MSE as the early stopping metric. The average test SEDIs ranged from 0.61 to 0.67  and the average test CSIs ranged from 0.42 to 0.5. When we used the SEDI as the early stopping metric, the average test SEDIs were consistently around 0.68 and the CSIs were around 0.49. In addition, when the undefined SEDIs are limited for calculating the average SEDI, exemplified by the one-month-ahead prediction, the SEDI highly correlates with the CSI. Because a high CSI indicates effective detection while considering both false alarms and misses, i.e. the precision and recall, the SEDI and CSI are typically high when the precision and recall are both high and closely matched. Therefore, we suggest more investigation into the selection of metrics for early stopping and evaluation.

\subsection{Prediction for Hotspot Locations}

We further examined the short-term forecasts at the 12 selected MHW hotspots \citep{oliver2021marine}, where key historical MHWs have been observed since 2000. The detailed results by location are in the Appendix. While the most effective approach varied by location, the ranking approach $m=25$ (7 out of 12) and the BMSE loss (10 out of 12) emerged as the most likely to yield the highest SEDI for MHW prediction at the hotspots. At this stage, the calculations of the loss function and evaluation metric for early stopping were averaged over all locations rather than being specific to certain nodes or areas. It could be assumed that if the calculations were made on a node or the average of a limited area, the performance for that node or area could potentially be improved and stabilized.

\subsection{Prediction for Longer Terms}

For long-lead forecasts, we first experimented with the sliding window with and without temporal diffusion (Table \ref{tb:temp}). We noticed that with regard to training time, using only the sliding window method resulted in the least time spent. Training with the diffusion process scaled with the lead time, i.e. the number of interpolators trained.
\begin{table*}[h!]
\centering \small
\begin{tabular}{llllll}
\toprule
Lead time                 & Method   & Precision   & Recall      & CSI         & SEDI        \\ \midrule
\multirow{3}{*}{1} & none     & 0.5921±0.14 & 0.7408±0.16 & 0.4961±0.14 & 0.6821±0.15 \\
                   & LSTM aggregator & 0.6047±0.15 & 0.704±0.16  & 0.4875±0.14 & 0.6648±0.15 \\
                   & diffusion   & 0.5921±0.14 & 0.7408±0.16 & 0.4961±0.14 & 0.6821±0.15 \\ \midrule
\multirow{3}{*}{2} & none     & 0.4684±0.16 & 0.6368±0.23 & 0.3692±0.15 & 0.4645±0.21 \\
                   & LSTM aggregator & 0.6517±0.27 & 0.1588±0.13 & 0.1468±0.12 & 0.3059±0.18 \\
                   & diffusion   & 0.4811±0.16 & 0.611±0.23  & 0.3726±0.16 & 0.4731±0.21 \\ \midrule
\multirow{3}{*}{3} & none     & 0.4275±0.17 & 0.5571±0.26 & 0.3182±0.16 & 0.3571±0.23 \\
                   & LSTM aggregator & 0.5682±0.31 & 0.0962±0.13 & 0.0835±0.1  & 0.2155±0.18 \\
                   & diffusion   & 0.4183±0.18 & 0.565±0.26  & 0.3147±0.16 & 0.3468±0.23 \\ \midrule
\multirow{3}{*}{4} & none     & 0.4064±0.2  & 0.4459±0.3  & 0.2534±0.16 & 0.2603±0.24 \\
                   & LSTM aggregator & 0.5353±0.39 & 0.0263±0.05 & 0.0241±0.05 & 0.1111±0.16 \\
                   & diffusion   & 0.3757±0.17 & 0.5866±0.28 & 0.2945±0.16 & 0.2703±0.24 \\ \midrule
\multirow{3}{*}{5} & none     & 0.4016±0.22 & 0.2993±0.26 & 0.1932±0.15 & 0.2042±0.23 \\
                   & LSTM aggregator & 0.484±0.39  & 0.0252±0.09 & 0.0182±0.05 & 0.0882±0.19 \\
                   & diffusion   & 0.4286±0.25 & 0.2282±0.25 & 0.1522±0.15 & 0.2025±0.21 \\ \midrule
\multirow{3}{*}{6} & none     & 0.4633±0.32 & 0.0693±0.18 & 0.0424±0.09 & 0.1482±0.19 \\
                   & LSTM aggregator & 0.4583±0.38 & 0.0311±0.12 & 0.019±0.05  & 0.1012±0.18 \\
                   & diffusion   & 0.4419±0.37 & 0.0216±0.12 & 0.0102±0.05 & 0.1391±0.21 \\ \bottomrule
\end{tabular}
\caption{Evaluation for short-to-long-lead MHW forecasts.}\label{tb:temp}
\end{table*}

We also experimented with three aggregators in GraphSAGE. The mean aggregator outperformed the pooling aggregator initially. With the LSTM aggregator, especially for long terms, the WMSE loss returned large average precisions for MHWs, with similar performance to the standard MSE. However, the effectiveness of the imbalanced losses was not substantial for longer-lead forecasts using the LSTM aggregator, and training with the LSTM aggregator was more than ten times slower than with the mean aggregator.

Generally, in terms of the CSI, whether or not the diffusion process was used, the results were similar. We noticed that for the forecasts longer than five months ahead, the SEDI could not be the primary evaluation metric because the index is undefined when either the precision or recall is zero, and nodes with undefined SEDIs were excluded from the computation of the average SEDI. However, the range of the CSI, precision, and recall is $[0,1]$ and all nodes were included in computing the average CSI. Hence, the six-month-ahead forecasts had average SEDIs around 0.14 but their average CSIs were almost zero, implying the unpredictability for six-month-ahead MHW forecasts so far. We also trained models for 12, 18, and 24 months, and they did not generate usable forecasts for nearly all nodes. Another set of experiments on different forecasting window sizes are described in the Appendix.

\begin{table*}[h!]
\centering \small
\begin{tabular}{llllll}
\toprule
Lead time                 & Method     & Precision   & Recall      & CSI         & SEDI        \\ \midrule
\multirow{2}{*}{2} & none       & 0.4881±0.16 & 0.615±0.24  & 0.3693±0.15 & 0.4783±0.21 \\
                   & diffusion  & 0.4843±0.16 & 0.6189±0.24 & 0.3692±0.15 & 0.4788±0.21 \\ \midrule
\multirow{2}{*}{3} & none       & 0.4666±0.2  & 0.4374±0.28 & 0.2762±0.17 & 0.3459±0.22 \\
                   & diffusion  & 0.4326±0.18 & 0.5342±0.28 & 0.3012±0.16 & 0.3521±0.23 \\ \midrule
\multirow{2}{*}{4} & none       & 0.4407±0.22 & 0.3548±0.29 & 0.2176±0.16 & 0.2665±0.23 \\
                   & diffusion  & 0.4093±0.19 & 0.3991±0.25 & 0.2449±0.15 & 0.2579±0.23 \\ \midrule
\multirow{2}{*}{5} & none       & 0.4391±0.26 & 0.2167±0.26 & 0.1405±0.15 & 0.2071±0.22 \\
                   & diffusion  & 0.3824±0.2  & 0.3668±0.29 & 0.2113±0.16 & 0.2037±0.23 \\ \midrule
\multirow{2}{*}{6} & none       & 0.4362±0.29 & 0.1358±0.21 & 0.091±0.12  & 0.1681±0.21 \\
                   & diffusion  & 0.3887±0.24 & 0.2231±0.25 & 0.14±0.14   & 0.1555±0.21 \\ \bottomrule
\end{tabular}
\caption{Evaluation without sliding windows for two-to-six-month-ahead MHW forecasts.}\label{tb:tempws1}
\end{table*}
\begin{figure*}[h!]
\centering
\includegraphics[scale=0.21]{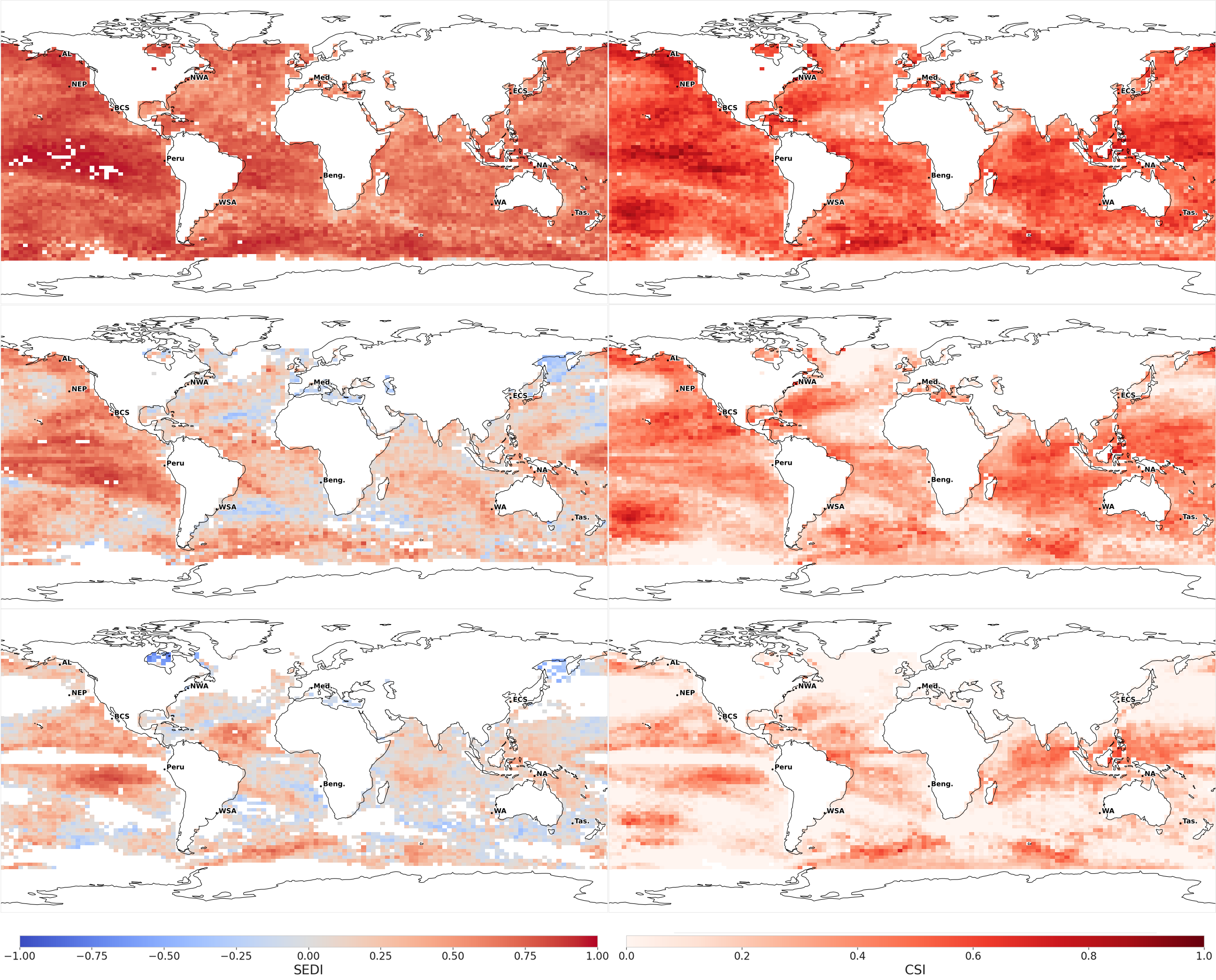}
\caption{The SEDI (left) and CSI (right) maps for MHW prediction: a one-month-ahead forecast with the BMSE (first line), a four-month-ahead forecast with the WMSE trained in temporal diffusion (second line), and a six-month-ahead forecast with the WMSE trained in temporal diffusion without sliding windows (third line). All used the $m=25$ graph construction method.}
\end{figure*}
We further investigated whether the diffusion process performed better without the sliding window (Table \ref{tb:tempws1}). When using just one time step inputs, whether or not the diffusion process was used returned similar average SEDIs and CSIs for two-month-ahead forecasts. However, when the number of lead months increased, although the SEDIs of the two methods remained similar, the gap between the CSIs grew larger, where the diffusion returned a larger average CSI. A larger average CSI indicates useful MHW forecasts at more locations. Also, the experiment further suggests that the SEDI, due to its undefined aspects, may not be as a robust evaluation metric as the CSI. In all, the results demonstrated that using one time step inputs and the diffusion process could generate generally better long-term MHW forecasts than conventional approaches using a long window.

Compared with the numerical model ensemble \citep{jacox2022global}, using a lead time of three to four months, both our models and the numerical models could provide higher MHW forecast skills evaluated by the SEDI in the equatorial Pacific, northwest Pacific near North America, south Pacific near South America, and equatorial Atlantic near South America than in other regions. Notably, our models also had higher SEDIs in the middle south Pacific, equatorial Atlantic near Africa, south Atlantic, and high-latitude Indian Ocean.

\section{Limitations and Future Work}

Due to the expansive scope of our study and computational resource constraints, we could not replicate model training for all configurations outlined in the Methodology Section, particularly for training with the LSTM aggregator and temporal diffusion process for forecasts beyond three months. We utilized the largest adjacency matrices that our time and resources allowed. The stability of our results across minor configuration adjustments, such as altering lead times, might indicate the models' consistency. Future investigations may concentrate on one aspect of this research and employ more computationally efficient GNN architectures to enable extensive robustness checks. While we have explored the perspectives of graph representation and imbalanced regression, the flexibility in graph construction and the variety of loss functions and evaluation metrics suggest that these areas warrant further exploration for climate extreme forecasts.

Future research could include exploring the flexibility of graph construction and other GNN architectures, targeting model training at MHW hotspots, refining imbalanced loss functions to balance the precision and recall for high SEDI or CSI values, optimizing the use of temporal diffusion, and adding new variables using domain knowledge.

\bibliography{aaai25}

\appendix

\section*{Appendix}

The Appendix includes experiment details, algorithms, additional experiment results, code accessibility, and data accessibility.

\subsection*{Experiment Details}

 For training the GraphSAGE models, the graph construction was further implemented using the Data object in PyTorch Geometric \citep{fey2019fast}. The attributes: ``x'' was the node feature matrix, ``y'' was the node label matrix, ``edge index'' was transformed from a selected adjacency matrix, ``edge attr'' was none, ``num nodes'' was the number of rows of the node feature matrix, ``num edges'' was the rank of the adjacency matrix, ``has isolated nodes'' was true, ``has self loops'' was false, and ``is undirected'' was true.

The GraphSAGE models were implemented in Python using PyTorch \citep{paszke2019pytorch} and PyTorch Geometric \citep{fey2019fast}. We deployed the following GraphSAGE structure: There are two SAGE convolutional layers, where The first layer receives the number of features equal to the window size as input and outputs 15 features, and the second layer receives 15 features and outputs one feature (the SSTA value for the target month). The optimizer is the AdamW, with a 0.01 learning rate and a 0.0001 weight decay. The activation function is the hyperbolic tangent. The loss function is one of the standard MSE, BMSE, and WMSE. The models predict SSTAs as a regression problem and then classifies the predicted SSTAs into binary classes: MHWs and non-MHWs. The MSE for SSTA predictions and the precision, recall, CSI, and SEDI for MHWs are reported. PyTorch Geometric provided three aggregation functions for SAGE convolutional layers and we used the default mean aggregation, otherwise specified.

Specifically for the temporal diffusion process, we employed an architecture analogue to \cite{cachay2023dyffusion}. There are the number of interpolator GraphSAGE models equal to the number of lead month less one, and there is one forecaster GraphSAGE. The GraphSAGE as an interpolator adds a dropout layer following each SAGE convolutional layer. The output of each interpolator is the SSTA values at one time step between the input time step(s) and the target time step. The number of input features to the forecaster is the number of features equal to the window size plus the number of interpolators (starting with zero before being interpolated). Assume the target lead time step is $m$ and the last input time step is $t$. For each training epoch, the forecaster is trained at the beginning to predict SSTAs at $m$. Then the first interpolator is trained to predict for $t+1$ and the forecaster is refined with input features and interpolated values at $t+1$. This process is repeated until the last interpolator for $m-1$ is trained and the forecaster is refined. It moves to the next training epoch. The dropout stochasticity is activated during training and disabled for validation or inference.

In training, we used early stopping in order to prevent overfitting associated with long training time. We set patience to be 40 and the maximum number of training epochs to be 200, besides patience of 30 and training epochs of 50 for the models using the temporal diffusion process or the LSTM aggregator, which cost long time in practice. The selection of the metric for early stopping is an open question, which can be either the loss metric or one evaluation metric on the test data \citep{terry2021statistically}. For example, in the traditional image classification, either the cross-entropy loss or accuracy can be the early stopping metric. In our preliminary experiments for tuning hyperparameters with one-month-ahead SSTA and MHW prediction, we found that using the SEDI as the early stopping metric could returned substantially larger average SEDIs and CSIs compared with using the MSE. Therefore, we used the SEDI as the metric to control early stopping. After each training epoch, the model configuration with the largest overall SEDI over the test data was saved and the training process continued for the number of epochs equal to the patience, until the next configuration with the largest test SEDI was found. If configurations with a largest SEDI continued to be found within epochs, the training was ended at the maximum epoch. To avoid potential large SEDIs returned by the randomness in early epochs, the early stopping strategy started at the sixth epoch.

The GraphSAGE models were trained on a shared Nvidia Tesla v100 machine. Therefore, there could be variations in training time caused by random GPU usage, so we did not record the training time. In practice, the differences in training time for using the LSTM aggregator, the diffusion process, and not using either were substantially different, which will be reported in the results.

\subsection*{Algorithms}

Algorithms \ref{ag:adjmat2}, \ref{ag:graphsage}, and \ref{alg:train_diffus}.

\begin{algorithm*}[h!]
\caption{Sorted Kendall rank correlation-based adjacency matrix construction.}\label{ag:adjmat2}
\begin{algorithmic}
\Require $node\_feat\_mat$ (Node feature matrix)
\Require $m$ (Number of edges per node)
\Ensure $adj\_mat$ (Generated adjacency matrix)
\Procedure{GetAdjMat}{$node\_feat\_mat, m$}
    \State $num\_nodes \gets \text{length of } node\_feat\_mat$
    \State Initialize $adj\_list$ as empty list
    \For{$i \gets 0$ \textbf{to} $num\_nodes-1$}
        \State Initialize $correlations$ as empty list
        \For{$j \gets 0$ \textbf{to} $num\_nodes-1$}
            \If{$i \neq j$}
                \State $correlation \gets kendall\_tau(node\_feat\_mat[i], node\_feat\_mat[j])[0]$
                \State Append $(j, correlation)$ to $correlations$
            \EndIf
        \EndFor
        \State $correlations \gets$ Sort $correlations$ by $-|\text{correlation}|$ and take first $m$
        \ForAll{$(j, \_)$ in $correlations$}
            \State Append $(i, j)$ and $(j, i)$ to $adj\_list$
        \EndFor
    \EndFor
    \State $adj\_list \gets$ Sort $adj\_list$ by first then second element
    \State $adj\_mat \gets$ Convert $adj\_list$ to a 2D NumPy array and transpose
    \State \Return $adj\_mat$
\EndProcedure
\end{algorithmic}
\end{algorithm*}

\begin{algorithm*}[h!]
\caption{GraphSAGE model for SST and SSTA forecasts.}\label{ag:graphsage}
\begin{algorithmic}
\Require $num\_graphs$ (Number of graphs)
\Ensure $x\_concat$ (Concatenated output)
\State $in\_channels \gets 12$
\State $hid\_channels \gets 15$
\State $out\_channels \gets 1$
\State $aggr \gets \text{`mean'}$
\State $convs \gets$ list of $num\_graphs$ SAGEConv sequential pairs
\Procedure{MultiGraphSage}{$in\_channels, hid\_channels, out\_channels, num\_graphs, aggr$}
    \For{$i \gets 0$ to $num\_graphs-1$}
        \State $convs[i] \gets$ Sequential(SAGEConv($in\_channels, hid\_channels, aggr$), 
        \Statex SAGEConv($hid\_channels, out\_channels, aggr$))
    \EndFor
\EndProcedure
\Function{Forward}{$data\_list$}
    \State Initialize $x\_list$ as empty list
    \ForAll{$data$ in $data\_list$}
        \State $x \gets data.x$
        \ForAll{$layer$ in $convs$}
            \State $x \gets layer(x, data.edge\_index)$
            \State $x \gets \tanh(x)$
        \EndFor
        \State Append $x$ to $x\_list$
    \EndFor
    \State $x\_concat \gets$ concatenate $x\_list$
    \State \Return $x\_concat$
\EndFunction
\end{algorithmic}
\end{algorithm*}

\begin{algorithm*}[h!]
\caption{Training GraphSAGE models using a temporal diffusion process.}\label{alg:train_diffus}
\begin{algorithmic}
\Require $num\_epochs$ (Number of training epochs)
\Require $lead\_time$ (Lead time for prediction)
\Require $train\_graph\_list\_fc$ (Training graphs for Forecaster)
\Require $train\_graph\_list\_ipt$ (Training graphs for Interpolator(s))
\Require $window\_size$ (Window size for features)
\For{$epoch \gets 1$ \textbf{to} $num\_epochs$}
    \State Initialize $pred\_node\_feat\_list\_fc$ as empty list
    \For{$i \gets 1$ \textbf{to} $lead\_time - 1$}
        \State Initialize $pred\_node\_feat\_list\_ipt$ as empty list
        \State Train/refine Forecaster
        \ForAll{$data$ in $train\_graph\_list\_fc$}
            \State Perform forecaster training steps
            \State Append $output$ from Forecaster to $pred\_node\_feat\_list\_fc$
        \EndFor
        \State Update $train\_graph\_list\_ipt$ with forecasted features
        \State Train one Interpolator for the current lead time
        \ForAll{$data$ in $train\_graph\_list\_ipt$}
            \State Perform interpolator training steps
            \State Append $output$ from Interpolator to $pred\_node\_feat\_list\_ipt$
        \EndFor
        \State Update $train\_graph\_list\_fc$ with interpolated features
    \EndFor
    \State Refine Forecaster for the last time in the current epoch
    \ForAll{$data$ in $train\_graph\_list\_fc$}
        \State Perform forecaster refinement steps
    \EndFor
\EndFor
\end{algorithmic}
\end{algorithm*}

\subsection*{Additional Experiment Results}

\begin{figure*}[h!]
\centering
\includegraphics[scale=0.375]{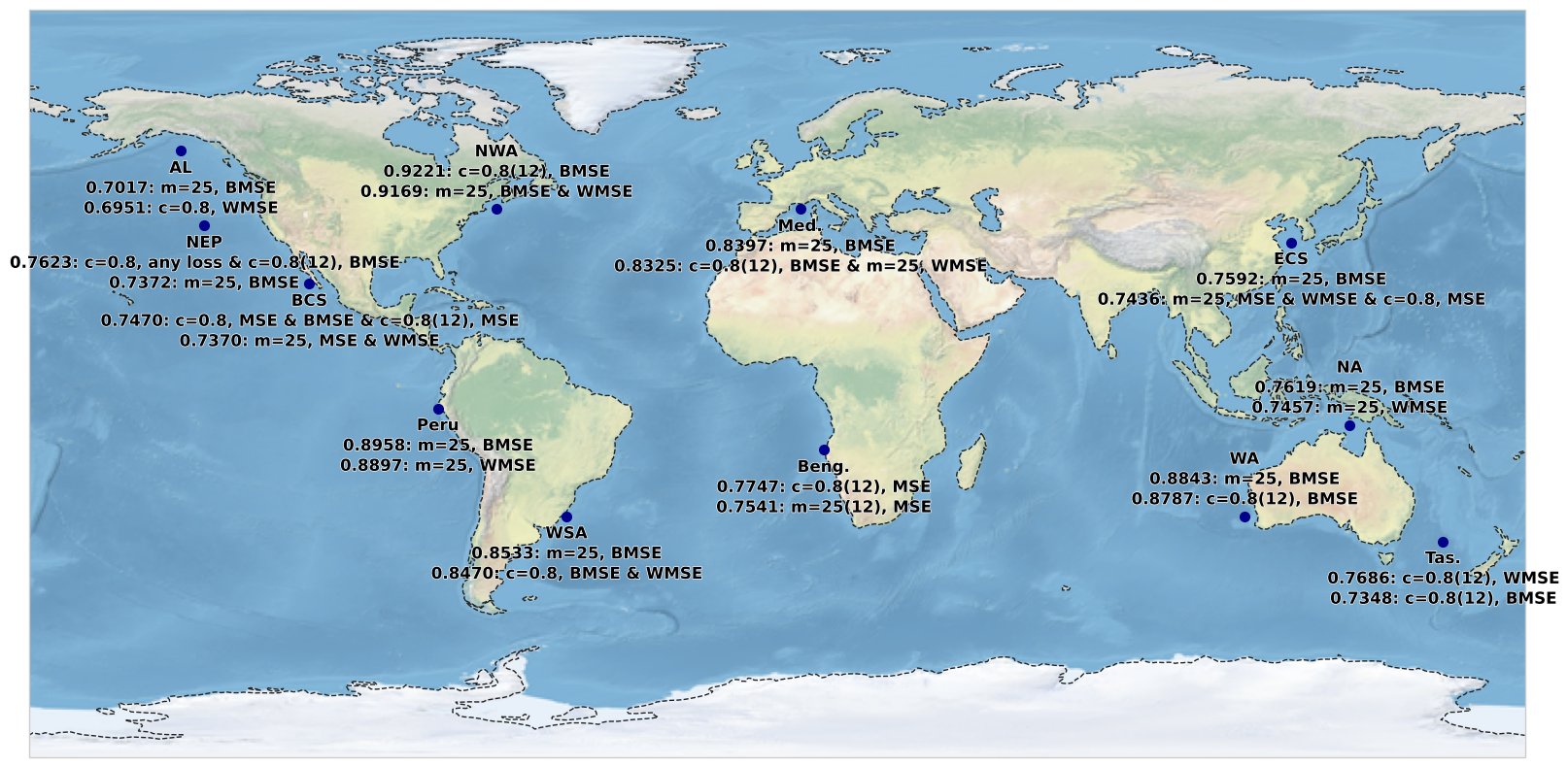}
\caption{The two highest SEDIs with the corresponding graph construction methods and loss functions at the 12 selected locations. Abbreviations are explained in the Appendix text.}\label{fg:locs_sedi}
\end{figure*}

Extended from the prediction for hotspot locations, Figure \ref{fg:locs_sedi} summarizes the two highest SEDIs for each location, along with the corresponding graph construction methods and loss functions.

Tables \ref{tb:al} to \ref{tb:wsa} detail the results for the 12 hotspot locations respectively. Abbreviations: AL, Gulf of Alaska and Bering Sea; Beng., Benguela; BCS, Baja California Sur; ECS, East China Sea; Med., Mediterranean Sea; NA, northern Australia; NEP, northeast Pacific; NWA, northwest Atlantic; Tas., Tasman Sea; WA, Western Australia; WSA, western South Atlantic.
\begin{table*}[h!]
\centering \small
\begin{tabular}{lllllll}
\toprule
Location             & Adjacency matrix                      & Loss & Precision & Recall & CSI    & SEDI   \\ \midrule
\multirow{12}{*}{AL} & \multirow{3}{*}{$c=0.8$}     & MSE  & 0.5484    & 0.6296 & 0.4146 & 0.6803 \\
                     &                              & BMSE & 0.5152    & 0.6296 & 0.3953 & 0.6591 \\
                     &                              & WMSE & 0.5294    & 0.6667 & 0.4186 & 0.6951 \\ \cline{2-7}
                     & \multirow{3}{*}{$c=0.8$, 12} & MSE  & 0.5313    & 0.6296 & 0.4048 & 0.6697 \\
                     &                              & BMSE & 0.5714    & 0.5926 & 0.4103 & 0.6671 \\
                     &                              & WMSE & 0.5714    & 0.5926 & 0.4103 & 0.6671 \\ \cline{2-7}
                     & \multirow{3}{*}{$m=25$}      & MSE  & 0.5484    & 0.6296 & 0.4146 & 0.6803 \\
                     &                              & BMSE & 0.5862    & 0.6296 & 0.4359 & 0.7017 \\
                     &                              & WMSE & 0.5667    & 0.6296 & 0.425  & 0.691 \\ \cline{2-7}
                     & \multirow{3}{*}{$m=25$, 12}  & MSE  & 0.5667    & 0.6296 & 0.425  & 0.691 \\
                     &                              & BMSE & 0.5517    & 0.5926 & 0.4    & 0.6556 \\
                     &                              & WMSE & 0.5714    & 0.5926 & 0.4103 & 0.6671 \\ \bottomrule
\end{tabular}
\caption{Evaluation for one-month-ahead MHW forecasts at AL.}\label{tb:al}
\end{table*}

\begin{table*}[h!]
\centering \small
\begin{tabular}{lllllll}
\toprule
Location             & Adjacency matrix                      & Loss & Precision & Recall & CSI    & SEDI   \\ \midrule
\multirow{12}{*}{Beng.} & \multirow{3}{*}{$c=0.8$}     & MSE  & 0.4583    & 0.6875 & 0.3793 & 0.7541 \\
                        &                              & BMSE & 0.4583    & 0.6875 & 0.3793 & 0.7541 \\
                        &                              & WMSE & 0.4583    & 0.6875 & 0.3793 & 0.7541 \\ \cline{2-7}
                        & \multirow{3}{*}{$c=0.8$, 12} & MSE  & 0.4138    & 0.75   & 0.3636 & 0.7747 \\
                        &                              & BMSE & 0.4783    & 0.6875 & 0.3929 & 0.7628 \\
                        &                              & WMSE & 0.4167    & 0.625  & 0.3333 & 0.6889 \\ \cline{2-7}
                        & \multirow{3}{*}{$m=25$}      & MSE  & 0.4583    & 0.6875 & 0.3793 & 0.7541 \\
                        &                              & BMSE & 0.5       & 0.6875 & 0.4074 & 0.7716 \\
                        &                              & WMSE & 0.4583    & 0.6875 & 0.3793 & 0.7541 \\ \cline{2-7}
                        & \multirow{3}{*}{$m=25$, 12}  & MSE  & 0.5       & 0.6875 & 0.4074 & 0.7716 \\
                        &                              & BMSE & 0.4348    & 0.625  & 0.3448 & 0.6988 \\
                        &                              & WMSE & 0.4583    & 0.6875 & 0.3793 & 0.7541 \\ \bottomrule
\end{tabular}
\caption{Evaluation for one-month-ahead MHW forecasts at Beng.}\label{tb:beng}
\end{table*}

\begin{table*}[h!]
\centering \small
\begin{tabular}{lllllll}
\toprule
Location             & Adjacency matrix                      & Loss & Precision & Recall & CSI    & SEDI   \\ \midrule
\multirow{12}{*}{BCS} & \multirow{3}{*}{$c=0.8$}     & MSE  & 0.5652    & 0.65   & 0.4333 & 0.747 \\
                      &                              & BMSE & 0.5652    & 0.65   & 0.4333 & 0.747 \\
                      &                              & WMSE & 0.5417    & 0.65   & 0.4194 & 0.737 \\ \cline{2-7}
                      & \multirow{3}{*}{$c=0.8$, 12} & MSE  & 0.5652    & 0.65   & 0.4333 & 0.747 \\
                      &                              & BMSE & 0.5217    & 0.6    & 0.3871 & 0.6928 \\
                      &                              & WMSE & 0.55      & 0.55   & 0.3793 & 0.6703 \\ \cline{2-7}
                      & \multirow{3}{*}{$m=25$}      & MSE  & 0.5217    & 0.6    & 0.3871 & 0.6928 \\
                      &                              & BMSE & 0.5417    & 0.65   & 0.4194 & 0.737 \\
                      &                              & WMSE & 0.5417    & 0.65   & 0.4194 & 0.737 \\ \cline{2-7}
                      & \multirow{3}{*}{$m=25$, 12}  & MSE  & 0.5455    & 0.6    & 0.4    & 0.7038 \\
                      &                              & BMSE & 0.5238    & 0.55   & 0.3667 & 0.6579 \\
                      &                              & WMSE & 0.5714    & 0.6    & 0.4138 & 0.715 \\ \bottomrule
\end{tabular}
\caption{Evaluation for one-month-ahead MHW forecasts at BCS.}\label{tb:bcs}
\end{table*}

\begin{table*}[h!]
\centering \small
\begin{tabular}{lllllll}
\toprule
Location             & Adjacency matrix                      & Loss & Precision & Recall & CSI    & SEDI   \\ \midrule
\multirow{12}{*}{ECS} & \multirow{3}{*}{$c=0.8$}     & MSE  & 0.3636    & 0.75   & 0.3243 & 0.7436 \\
                      &                              & BMSE & 0.3529    & 0.75   & 0.3158 & 0.7358 \\
                      &                              & WMSE & 0.3529    & 0.75   & 0.3158 & 0.7358 \\ \cline{2-7}
                      & \multirow{3}{*}{$c=0.8$, 12} & MSE  & 0.2545    & 0.875  & 0.2456 & 0.7196 \\
                      &                              & BMSE & 0.3529    & 0.75   & 0.3158 & 0.7358 \\
                      &                              & WMSE & 0.3529    & 0.75   & 0.3158 & 0.7358 \\ \cline{2-7}
                      & \multirow{3}{*}{$m=25$}      & MSE  & 0.3636    & 0.75   & 0.3243 & 0.7436 \\
                      &                              & BMSE & 0.3871    & 0.75   & 0.3429 & 0.7592 \\
                      &                              & WMSE & 0.3636    & 0.75   & 0.3243 & 0.7436 \\ \cline{2-7}
                      & \multirow{3}{*}{$m=25$, 12}  & MSE  & 0.3333    & 0.6875 & 0.2895 & 0.6754 \\
                      &                              & BMSE & 0.3243    & 0.75   & 0.2927 & 0.7119 \\
                      &                              & WMSE & 0.3077    & 0.75   & 0.2791 & 0.6957 \\ \bottomrule
\end{tabular}
\caption{Evaluation for one-month-ahead MHW forecasts at ECS.}\label{tb:ecs}
\end{table*}

\begin{table*}[h!]
\centering \small
\begin{tabular}{lllllll}
\toprule
Location             & Adjacency matrix                      & Loss & Precision & Recall & CSI    & SEDI   \\ \midrule
\multirow{12}{*}{Med.} & \multirow{3}{*}{$c=0.8$}     & MSE  & 0.2857    & 0.5    & 0.2222 & 0.6212 \\
                       &                              & BMSE & 0.3529    & 0.75   & 0.3158 & 0.8253 \\
                       &                              & WMSE & 0.3125    & 0.625  & 0.2632 & 0.7255 \\ \cline{2-7}
                       & \multirow{3}{*}{$c=0.8$, 12} & MSE  & 0.2857    & 0.75   & 0.2609 & 0.7967 \\
                       &                              & BMSE & 0.375     & 0.75   & 0.3333 & 0.8325 \\
                       &                              & WMSE & 0.2667    & 0.5    & 0.2105 & 0.6088 \\ \cline{2-7}
                       & \multirow{3}{*}{$m=25$}      & MSE  & 0.3529    & 0.75   & 0.3158 & 0.8253 \\
                       &                              & BMSE & 0.4       & 0.75   & 0.3529 & 0.8397 \\
                       &                              & WMSE & 0.375     & 0.75   & 0.3333 & 0.8325 \\ \cline{2-7}
                       & \multirow{3}{*}{$m=25$, 12}  & MSE  & 0.3529    & 0.75   & 0.3158 & 0.8253 \\
                       &                              & BMSE & 0.3125    & 0.625  & 0.2632 & 0.7255 \\
                       &                              & WMSE & 0.3125    & 0.625  & 0.2632 & 0.7255 \\ \bottomrule
\end{tabular}
\caption{Evaluation for one-month-ahead MHW forecasts at Med.}\label{tb:med}
\end{table*}

\begin{table*}[h!]
\centering \small
\begin{tabular}{lllllll}
\toprule
Location             & Adjacency matrix                      & Loss & Precision & Recall & CSI    & SEDI   \\ \midrule
\multirow{12}{*}{NA} & \multirow{3}{*}{$c=0.8$}     & MSE  & 0.3077    & 0.5714 & 0.25   & 0.5974 \\
                     &                              & BMSE & 0.3448    & 0.7143 & 0.303  & 0.7294 \\
                     &                              & WMSE & 0.3448    & 0.7143 & 0.303  & 0.7294 \\ \cline{2-7}
                     & \multirow{3}{*}{$c=0.8$, 12} & MSE  & 0.22      & 0.7857 & 0.2075 & 0.6398 \\
                     &                              & BMSE & 0.3571    & 0.7143 & 0.3125 & 0.7375 \\
                     &                              & WMSE & 0.2963    & 0.5714 & 0.2424 & 0.587 \\ \cline{2-7}
                     & \multirow{3}{*}{$m=25$}      & MSE  & 0.3333    & 0.7143 & 0.2941 & 0.7212 \\
                     &                              & BMSE & 0.4       & 0.7143 & 0.3448 & 0.7619 \\
                     &                              & WMSE & 0.3704    & 0.7143 & 0.3226 & 0.7457 \\ \cline{2-7}
                     & \multirow{3}{*}{$m=25$, 12}  & MSE  & 0.3333    & 0.6429 & 0.2813 & 0.6702 \\
                     &                              & BMSE & 0.2963    & 0.5714 & 0.2424 & 0.587 \\
                     &                              & WMSE & 0.3333    & 0.6429 & 0.2813 & 0.6702 \\ \bottomrule
\end{tabular}
\caption{Evaluation for one-month-ahead MHW forecasts at NA.}\label{tb:na}
\end{table*}

\begin{table*}[h!]
\centering \small
\begin{tabular}{lllllll}
\toprule
Location             & Adjacency matrix                      & Loss & Precision & Recall & CSI    & SEDI   \\ \midrule
\multirow{12}{*}{NEP} & \multirow{3}{*}{$c=0.8$}     & MSE  & 0.5455 & 0.6667 & 0.4286 & 0.7623 \\
                      &                              & BMSE & 0.5455 & 0.6667 & 0.4286 & 0.7623 \\
                      &                              & WMSE & 0.5455 & 0.6667 & 0.4286 & 0.7623 \\ \cline{2-7}
                      & \multirow{3}{*}{$c=0.8$, 12} & MSE  & 0.5    & 0.6111 & 0.3793 & 0.7047 \\
                      &                              & BMSE & 0.5455 & 0.6667 & 0.4286 & 0.7623 \\
                      &                              & WMSE & 0.5    & 0.5    & 0.3333 & 0.6247 \\ \cline{2-7}
                      & \multirow{3}{*}{$m=25$}      & MSE  & 0.5    & 0.6111 & 0.3793 & 0.7047 \\
                      &                              & BMSE & 0.5789 & 0.6111 & 0.4231 & 0.7372 \\
                      &                              & WMSE & 0.5    & 0.6111 & 0.3793 & 0.7047 \\ \cline{2-7}
                      & \multirow{3}{*}{$m=25$, 12}  & MSE  & 0.5263 & 0.5556 & 0.3704 & 0.6772 \\
                      &                              & BMSE & 0.4737 & 0.5    & 0.3214 & 0.6113 \\
                      &                              & WMSE & 0.5263 & 0.5556 & 0.3704 & 0.6772 \\ \bottomrule
\end{tabular}
\caption{Evaluation for one-month-ahead MHW forecasts at NEP.}\label{tb:nep}
\end{table*}

\begin{table*}[h!]
\centering \small
\begin{tabular}{lllllll}
\toprule
Location              & Adjacency matrix                      & Loss & Precision & Recall & CSI    & SEDI   \\ \midrule
\multirow{12}{*}{NWA} & \multirow{3}{*}{$c=0.8$}     & MSE  & 0.5714 & 0.8    & 0.5    & 0.8702 \\
                      &                              & BMSE & 0.5455 & 0.8    & 0.48   & 0.8636 \\
                      &                              & WMSE & 0.5455 & 0.8    & 0.48   & 0.8636 \\ \cline{2-7}
                      & \multirow{3}{*}{$c=0.8$, 12} & MSE  & 0.5789 & 0.7333 & 0.4783 & 0.8323 \\
                      &                              & BMSE & 0.65   & 0.8667 & 0.5909 & 0.9221 \\
                      &                              & WMSE & 0.6111 & 0.7333 & 0.5    & 0.8405 \\ \cline{2-7}
                      & \multirow{3}{*}{$m=25$}      & MSE  & 0.5652 & 0.8667 & 0.52   & 0.9063 \\
                      &                              & BMSE & 0.619  & 0.8667 & 0.5652 & 0.9169 \\
                      &                              & WMSE & 0.5909 & 0.8667 & 0.5417 & 0.9116 \\ \cline{2-7}
                      & \multirow{3}{*}{$m=25$, 12}  & MSE  & 0.6    & 0.8    & 0.5217 & 0.8768 \\
                      &                              & BMSE & 0.6111 & 0.7333 & 0.5    & 0.8405 \\
                      &                              & WMSE & 0.6111 & 0.7333 & 0.5    & 0.8405 \\ \bottomrule
\end{tabular}
\caption{Evaluation for one-month-ahead MHW forecasts at NWA.}\label{tb:nwa}
\end{table*}

\begin{table*}[h!]
\centering \small
\begin{tabular}{lllllll}
\toprule
Location             & Adjacency matrix                      & Loss & Precision & Recall & CSI    & SEDI   \\ \midrule
\multirow{12}{*}{Peru} & \multirow{3}{*}{$c=0.8$}     & MSE  & 0.45   & 0.8182 & 0.4091 & 0.8715 \\
                       &                              & BMSE & 0.4286 & 0.8182 & 0.3913 & 0.8654 \\
                       &                              & WMSE & 0.4286 & 0.8182 & 0.3913 & 0.8654 \\ \cline{2-7}
                       & \multirow{3}{*}{$c=0.8$, 12} & MSE  & 0.4211 & 0.7273 & 0.3636 & 0.8064 \\
                       &                              & BMSE & 0.4444 & 0.7273 & 0.381  & 0.8142 \\
                       &                              & WMSE & 0.4667 & 0.6364 & 0.3684 & 0.7623 \\ \cline{2-7}
                       & \multirow{3}{*}{$m=25$}      & MSE  & 0.4706 & 0.7273 & 0.4    & 0.8221 \\
                       &                              & BMSE & 0.5625 & 0.8182 & 0.5    & 0.8958 \\
                       &                              & WMSE & 0.5294 & 0.8182 & 0.4737 & 0.8897 \\ \cline{2-7}
                       & \multirow{3}{*}{$m=25$, 12}  & MSE  & 0.4375 & 0.6364 & 0.35   & 0.7523 \\
                       &                              & BMSE & 0.5    & 0.6364 & 0.3889 & 0.7726 \\
                       &                              & WMSE & 0.4667 & 0.6364 & 0.3684 & 0.7623 \\ \bottomrule
\end{tabular}
\caption{Evaluation for one-month-ahead MHW forecasts at Peru.}\label{tb:peru}
\end{table*}

\begin{table*}[h!]
\centering \small
\begin{tabular}{lllllll}
\toprule
Location             & Adjacency matrix                      & Loss & Precision & Recall & CSI    & SEDI   \\ \midrule
\multirow{12}{*}{Tas.} & \multirow{3}{*}{$c=0.8$}     & MSE  & 0.44   & 0.6471 & 0.3548 & 0.7083 \\
                       &                              & BMSE & 0.4    & 0.7059 & 0.3429 & 0.7263 \\
                       &                              & WMSE & 0.3871 & 0.7059 & 0.3333 & 0.7178 \\ \cline{2-7}
                       & \multirow{3}{*}{$c=0.8$, 12} & MSE  & 0.2766 & 0.7647 & 0.2549 & 0.651  \\
                       &                              & BMSE & 0.4138 & 0.7059 & 0.3529 & 0.7348 \\
                       &                              & WMSE & 0.48   & 0.7059 & 0.4    & 0.7686 \\ \cline{2-7}
                       & \multirow{3}{*}{$m=25$}      & MSE  & 0.3667 & 0.6471 & 0.3056 & 0.6609 \\
                       &                              & BMSE & 0.4348 & 0.5882 & 0.3333 & 0.6632 \\
                       &                              & WMSE & 0.4167 & 0.5882 & 0.3226 & 0.6526 \\ \cline{2-7}
                       & \multirow{3}{*}{$m=25$, 12}  & MSE  & 0.3243 & 0.7059 & 0.2857 & 0.6661 \\
                       &                              & BMSE & 0.4074 & 0.6471 & 0.3333 & 0.6893 \\
                       &                              & WMSE & 0.4    & 0.7059 & 0.3429 & 0.7263 \\ \bottomrule
\end{tabular}
\caption{Evaluation for one-month-ahead MHW forecasts at Tas.}\label{tb:tas}
\end{table*}

\begin{table*}[h!]
\centering \small
\begin{tabular}{lllllll}
\toprule
Location             & Adjacency matrix                      & Loss & Precision & Recall & CSI    & SEDI   \\ \midrule
\multirow{12}{*}{WA} & \multirow{3}{*}{$c=0.8$}     & MSE  & 0.44   & 0.7857 & 0.3929 & 0.827  \\
                     &                              & BMSE & 0.4138 & 0.8571 & 0.3871 & 0.8615 \\
                     &                              & WMSE & 0.4138 & 0.8571 & 0.3871 & 0.8615 \\ \cline{2-7}
                     & \multirow{3}{*}{$c=0.8$, 12} & MSE  & 0.4286 & 0.8571 & 0.4    & 0.8673 \\
                     &                              & BMSE & 0.4615 & 0.8571 & 0.4286 & 0.8787 \\
                     &                              & WMSE & 0.4444 & 0.8571 & 0.4138 & 0.873  \\ \cline{2-7}
                     & \multirow{3}{*}{$m=25$}      & MSE  & 0.3871 & 0.8571 & 0.3636 & 0.8498 \\
                     &                              & BMSE & 0.48   & 0.8571 & 0.4444 & 0.8843 \\
                     &                              & WMSE & 0.44   & 0.7857 & 0.3929 & 0.827  \\ \cline{2-7}
                     & \multirow{3}{*}{$m=25$, 12}  & MSE  & 0.4583 & 0.7857 & 0.4074 & 0.8339 \\
                     &                              & BMSE & 0.44   & 0.7857 & 0.3929 & 0.827  \\
                     &                              & WMSE & 0.44   & 0.7857 & 0.3929 & 0.827  \\ \bottomrule
\end{tabular}
\caption{Evaluation for one-month-ahead MHW forecasts at WA.}\label{tb:wa}
\end{table*}

\begin{table*}[h!]
\centering \small
\begin{tabular}{lllllll}
\toprule
Location             & Adjacency matrix                      & Loss & Precision & Recall & CSI    & SEDI   \\ \midrule
\multirow{12}{*}{WSA} & \multirow{3}{*}{$c=0.8$}     & MSE  & 0.3333 & 0.6    & 0.2727 & 0.6917 \\
                      &                              & BMSE & 0.381  & 0.8    & 0.3478 & 0.847  \\
                      &                              & WMSE & 0.381  & 0.8    & 0.3478 & 0.847  \\ \cline{2-7}
                      & \multirow{3}{*}{$c=0.8$, 12} & MSE  & 0.3043 & 0.7    & 0.2692 & 0.7453 \\
                      &                              & BMSE & 0.35   & 0.7    & 0.3043 & 0.7696 \\
                      &                              & WMSE & 0.2632 & 0.5    & 0.2083 & 0.5709 \\ \cline{2-7}
                      & \multirow{3}{*}{$m=25$}      & MSE  & 0.3333 & 0.7    & 0.2917 & 0.7615 \\
                      &                              & BMSE & 0.4    & 0.8    & 0.3636 & 0.8533 \\
                      &                              & WMSE & 0.3636 & 0.8    & 0.3333 & 0.8406 \\ \cline{2-7}
                      & \multirow{3}{*}{$m=25$, 12}  & MSE  & 0.3158 & 0.6    & 0.2609 & 0.6816 \\
                      &                              & BMSE & 0.3    & 0.6    & 0.25   & 0.6717 \\
                      &                              & WMSE & 0.3158 & 0.6    & 0.2609 & 0.6816 \\ \bottomrule
\end{tabular}
\caption{Evaluation for one-month-ahead MHW forecasts at WSA.}\label{tb:wsa}
\end{table*}

\begin{table*}[t]
\centering \small
\begin{tabular}{lllllll}
\toprule
Lead time                 & Window size            & Method    & Precision   & Recall      & CSI         & SEDI        \\ \midrule
\multirow{8}{*}{3} & \multirow{2}{*}{1}  & none      & 0.4666±0.2  & 0.4374±0.28 & 0.2762±0.17 & 0.3459±0.22 \\
                   &                     & diffusion & 0.4326±0.18 & 0.5342±0.28 & 0.3012±0.16 & 0.3521±0.23 \\ \cline{2-7}
                   & \multirow{2}{*}{3}  & none      & 0.4668±0.2  & 0.4213±0.28 & 0.2657±0.17 & 0.3334±0.23 \\
                   &                     & diffusion & 0.4402±0.18 & 0.4796±0.26 & 0.2916±0.16 & 0.3424±0.23 \\ \cline{2-7}
                   & \multirow{2}{*}{6}  & none      & 0.4887±0.22 & 0.3447±0.26 & 0.236±0.17  & 0.3168±0.22 \\
                   &                     & diffusion & 0.4416±0.19 & 0.4781±0.28 & 0.2871±0.16 & 0.3402±0.23 \\ \cline{2-7}
                   & \multirow{2}{*}{12} & none      & 0.4275±0.17 & 0.5571±0.26 & 0.3182±0.16 & 0.3571±0.23 \\
                   &                     & diffusion & 0.4183±0.18 & 0.565±0.26  & 0.3147±0.16 & 0.3468±0.23 \\ \hline
\multirow{8}{*}{6} & \multirow{2}{*}{1}  & none      & 0.4362±0.29 & 0.1358±0.21 & 0.091±0.12  & 0.1681±0.21 \\
                   &                     & diffusion & 0.3887±0.24 & 0.2231±0.25 & 0.14±0.14   & 0.1555±0.21 \\ \cline{2-7}
                   & \multirow{2}{*}{3}  & none      & 0.4725±0.34 & 0.0669±0.16 & 0.0453±0.09 & 0.1484±0.19 \\
                   &                     & diffusion & 0.2147±0.18 & 0.0123±0.1  & 0.0042±0.03 & 0.1377±0.23 \\ \cline{2-7}
                   & \multirow{2}{*}{6}  & none      & 0.2784±0.25 & 0.0154±0.11 & 0.0053±0.04 & 0.1595±0.22 \\
                   &                     & diffusion & 0.1918±0.34 & 0.0008±0.01 & 0.0007±0.01 & 0.1273±0.13 \\ \cline{2-7}
                   & \multirow{2}{*}{12} & none      & 0.4633±0.32 & 0.0693±0.18 & 0.0424±0.09 & 0.1482±0.19 \\
                   &                     & diffusion & 0.4419±0.37 & 0.0216±0.12 & 0.0102±0.05 & 0.1391±0.21  \\ \bottomrule
\end{tabular}
\caption{Evaluation with different window sizes for three-and-six-month-ahead MHW forecasts.}\label{tb:tempws}
\end{table*}

We experimented with different forecasting window sizes (Table \ref{tb:tempws}). The previous study demonstrated that when using a GraphSAGE model and a recursive model for long-term SST forecasts, a long window performed distinctly better than a short window in terms of the RMSE, despite increased inputs \citep{ning2024harnessing}. Notably on the contrary, for long-term MHW forecasts, a short window provided better results than a larger window with regard to the SEDI, and the diffusion process generally returned better performance than the models without it. The reason could be SST forecasts relies heavily on seasonality and annual cycles provided by a large window size, while for SSTA and MHW forecasts, observations far prior to the target time step do not provide useful information, and the one adjacent time step is highly correlated and sufficient for prediction. Also, simply using one time step reduces the amount of inputs.

\subsection*{Code Accessibility}

The package we developed to generate the arrays is available at \url{https://anonymous.4open.science/r/converter-4BE8}. The scripts for the models in this paper and other auxiliary functions are accessible at \url{https://anonymous.4open.science/r/mhw-D435}.

\subsection*{Data Accessibility}

The original ERA5 data can be downloaded from the Copernicus Climate Change Service,  \url{https://cds.climate.copernicus.eu/}, by selecting the ``ERA5 monthly averaged data on single levels from 1940 to present'' dataset. The generated SSTA graphs from the preprocessed ERA5, including the graphs from \cite{ning2024harnessing}, are accessible at \url{https://www.dropbox.com/scl/fo/brzad7hy7h55hq6cxi1e9/ANbHz8mpw-C0zPI7xxBsb3k?rlkey=80kbz4nbjhqltw40gvgvi8rdg&st=zaigim8c&dl=0}.

\end{document}